\documentclass[useAMS,usenatbib]{mn2e}

\title[Anomalous precession of planets on a Weyl conformastatic solution]{Anomalous precession of planets on a Weyl conformastatic solution}
\author[Abra\~{a}o J. S. Capistrano, Joice A. M. Penagos and Manuel S. Al\'{a}rcon]
{Abra\~{a}o J. S. Capistrano$^{1}$\thanks{E-mail:abraao.capistrano@unila.edu.br},
Joice A. M. Penagos $^{2}$\thanks{E-mail:bioastro@gmail.com}, Manuel S. Al\'{a}rcon$^{3}$\thanks{E-mail:mf.sanchez17@uniandes.edu.co}\\
$^{1,2}$Federal University of Latin-American Integration, \\
Technological Park of Itaipu, 85867-670, P.o.b: 2123, Foz do igua\c{c}u-PR, Brazil\\
$^{1}$Casimiro Montenegro Filho Astronomy Center,\\
Technological Park of Itaipu, 85867-900, Foz do igua\c{c}u-PR, Brazil\\
$^{3}$Department of Physics, Universidad de Los Andes, Bogot\'{a}, Colombia}

\begin{document}

\date{}

\pagerange{\pageref{firstpage}--\pageref{lastpage}} \pubyear{2016}

\maketitle

\label{firstpage}

\begin{abstract}
In this paper, we investigate the anomalous planets precession in the nearly-newtonian gravitational regime. This limit is obtained from the application of the slow motion condition to the geodesic equations without altering the geodesic deviation equations. Using a non-standard expression for the perihelion advance from the Weyl conformastatic vacuum solution as a model, we can describe the anomaly in planets precession compared with different observational data from Ephemerides of the Planets and the Moon (EPM2008 and EPM2011) and Planetary and Lunar Ephemeris (INPOP10a). As a result, using the Levenberg-Marquardt algorithm and calculating the related Chi-squared statistic, we find that the anomaly is statistical irrelevant in accordance with INPOP10a observations.  As a complement to this work, we also do an application to the relativistic precession of giant planets using observational data calibrated with the EPM2011.
\end{abstract}


\section{Introduction}

In a previous communication \citep{Capistrano}, we studied the perihelion advance for inner planets and comets obtaining a good agreement with observations. To this matter, we did not use nor the \emph{Parameterized Post-Newtonian approximation}(PPN)\citep{Nobili} neither related methods. Rather, we investigated the possibility of applications of a new gravitational regime, the Nearly Newtonian Approximation (NNA)\citep{wheeler}. This regime is obtained from the imposition of the slow motion condition ($v<<c$) on the geodesic equations, while the deviation equations are left alone. The novelty of the method resides in the influence of non linearities on the solutions at certain level making the resulting gravitational potential weaker that one from general relativity (GR) but stronger that the newtonian one. Another important aspect is that the diffeomorphic transformations from GR are broken, which allows at first to test NNA method to astrophysics problems.

We focus the present work on the description of the anomalous apsidal precession of planets. The effect of this precession is very tiny of order of -6 miliarcsec as identified by E.V.Pitjeva \citep{Pitjeva1,lorio1} based on the analysis of normal points observations of Cassini spacecraft (2004-2006) with EPM2008 ephemerides that suggested a retrograde precession of Saturn with relevant nonzero statistical significance, and, apparently, it could not be explained by standard theory of gravity, i.e, nor in newtonian regime neither in GR. The same situation might be applied to the other planets. Due to the smallness of the effect, high margin of errors and the powers of accuracy, the current debate lies on the quest of an ultimate conclusion on the confirmation or not of this effect. If so, an underlying physics must be taken into account in a quest of finding explanations of this subtle effect of gravity and on the understanding of gravity itself in a general manner. In addition, several proposals have been made from deviations of standard classical newtonian and Einstein's gravity \citep{Acedo} up to modifications of gravity itself, e.g, including (or not) dark matter \citep{lorio1,lorio3,lorio4,lue, khrip,Sereno}. For a review on this subject see \citep{lorio14} and references therein.

In this work, we use the same rationalization to test this NNA method in solar scale system in order to look for subtle non-linear effects on solar system gravity. In the second section, we review the main results of Weyl's vacuum conformastatic solution for a thin disk as shown in \cite{Capistrano}. On the other hand, differently from what we present in this work, the solutions for relativistic disk for a matter distribution can be found in, e.g, \citep{antonio,antonio2}. As a complement, we determine the perihelion precession of giant planets and compare to known results in literature. It is important to note that in the recent years there have been a renewed interest in the perihelion advance as one of fundamental test for any candidate to a gravitational theory. We testified a plethora of methods and new theories such as the modification of Newtonian Dynamics (MOND)\citep{schimit}, azimuthally symmetric theory of gravitation based on the study of Poisson equation \citep{nambuya}, higher dimensional theories such as Kaluza-Klein five-dimensional gravity \citep{wesson}, Yukawa-like Modified Gravity \citep{lorenzo}, Horava-Lifshitz gravity \citep{harko}, brane-world and variants \citep{mak,cheung,sumanta,sepangi,lorio1,lorio2} and approaches in the weak field/slow motion limits \citep{Avalos,Arakida,Deliseo,Deng,Feldman,Kalinowski,lorio3,lorio4,lorio5,lorio6,lorio7,lorio8,lorio9,lorio10,lorio11,lorio12,lorio13,Ruggiero,Xie,Wilhelm}.

In the third section, we investigate the possibility to apply the NNA method to the anomalous precession of the planets comparing a prior heuristic approach with Chi-squared statistic. To this matter, we compare our results with different observations on Ephemerides of the Planets and the Moon (EPM2008 \citep{Pitjeva1}, EMP2011 \citep{Pitjeva,Pitjev}) and on Planetary and Lunar Ephemeris (INPOP10a)\citep{Fienga}. In the final section, we present final remarks and future prospects.

\section{The Weyl disk conformastatic solution}
\begin{table*}
 \centering
 \begin{minipage}{140mm}
  \caption{Relevant elements for computing the perihelion precession. The following data below can be found in Nasa Planetary Fact Sheets database (http://nssdc.gsfc.nasa.gov/planetary/planetfact.html). The semi-major axis is shown in astronomical units ($1 AU =  1.49598 \times10^{11} m$). The orbital periods are in units of years.}
  \begin{tabular}{@{}llrrrrlrlr@{}}
  \hline
   Object        &   Semi-major axis ($\times 10^{6}km$)       & eccentricity &  Period (yr)   \\
  \hline
  Jupiter        & $\;\;\;$  778.57                            & 0.0489                   &  11.86    \\
  Saturn         &  $\;\;\;$ 1,433.53                          & 0.0565                   &  29.46    \\
  Uranus         &  $\;\;\;$ 2,872.46                          & 0.0457                   &  84.01    \\
  Neptune        &  $\;\;\;$ 4,495.06                          & 0.0113                   &  164.79    \\
  \hline
\end{tabular}
\end{minipage}
\end{table*}

The Weyl's conformastatic solution is obtained from Weyl's line element \citep{weyl}
\begin{equation}
ds^{2}=e^{2\left(\lambda-\sigma\right)}dr^{2}+r^{2}e^{-2\sigma}d\theta^{2}+e^{2(\lambda-\sigma)}dz^{2}-e^{2\sigma}dt^{2}\;,
\end{equation}
where the coefficients of the metric are defined as $\lambda=\lambda(r,z)$ and $\sigma=\sigma(r,z)$. The exterior
gravitational field in the cylinder outskirts is given by Einstein's
vacuum equations
 \begin{eqnarray}
 && - \lambda_{,r} + r\sigma_{,r}^{2} -r
\sigma_{,z}^{2} =0\;,
\label{eq:first}\\
 &&\sigma_{,r}+r\sigma_{,rr}+r\sigma_{,zz}=0\;, \label{eq:second}\\
&& 2r\sigma_{,r}\sigma_{,z}
=\lambda_{,z}\;,\label{eq:fourth}
\end{eqnarray}
where the terms $(,r)$, $(,z)$ and $(,,r)$, $(,,z)$ denote, respectively, the first and the second derivatives with respect to the variables $r$ and $z$.
As shown in Weyl's original paper, the cylinder solution is diffeomorphic to a Schwarzschild's solution, hence the metric remains asymptotically flat \citep{weyl,rosen,zipoy,gau,steph}, which can be useful to astrophysics purposes.

In order to model the movement of an orbiting body in a near circular orbit, we assume that the cylinder thickness $h_{0}$ is smaller than its radius
$R_{0}$, i.e., $h_{0}<< R_{0}$. As a consequence, one can expand the coefficients $\lambda(r,z)$ and $\sigma(r,z)$ into a MacLaurin's series such that
\begin{equation}\label{eq23}
\sigma (r,z) \approx \sigma (r,0) + z{\left. {{{\partial \sigma (r,z)} \over {\partial z}}} \right|_{z = 0}} + {z^2}{\left. {{{{\partial ^2}\sigma (r,z)} \over {\partial {z^2}}}} \right|_{z = 0}} +  \cdots,
\end{equation}

\begin{equation}\label{eq24}
\lambda (r,z) \approx \lambda (r,0) + z{\left. {{{\partial \lambda (r,z)} \over {\partial z}}} \right|_{z = 0}} + {z^2}{\left. {{{{\partial ^2}\lambda (r,z)} \over {\partial {z^2}}}} \right|_{z = 0}} +  \cdots.
\end{equation}
and can be truncated in second order \citep{Capistrano}. Hence, one can define
\begin{equation}\label{eq25}
  \sigma(r,z) = A(r) + a(r)z + c(r){z^2},
\end{equation}
where we denote $A(r)=\sigma (r,0)$, $a(r)= {\left. {{{\partial \sigma (r,z)} \over {\partial z}}} \right|_{z = 0}}$ and $c(r)=\left.{{{{\partial ^2}\sigma (r,z)} \over {\partial {z^2}}}} \right|_{z = 0}$. In addition, the coefficient $\lambda (r,z)$ can be define the same as
\begin{equation}\label{eq25a}
  \lambda (r,z) = B(r) + b(r)z + d(r){z^2},  	
\end{equation}
where we denote $B(r)=\lambda (r,0)$, $b(r)= {\left. {{{\partial \lambda (r,z)} \over {\partial z}}} \right|_{z = 0}}$ and $d(r)=\left.{{{{\partial ^2}\lambda (r,z)} \over {\partial {z^2}}}} \right|_{z = 0}$.

After using eqs.(\ref{eq25}) and (\ref{eq25a}) and solving the non-linear system in eqs.(\ref{eq:first}), (\ref{eq:second})and (\ref{eq:fourth}), one can find the coefficients $\sigma(r,z)$ and $\lambda(r,z)$
\begin{equation}\label{eq42}
{\sigma (r,z) = {{{k_0}} \over 2}\ln (r) - {{{{c_0}{r^2}} \over 2}} + {a_0}z + {c_0}{z^2}}\;,
\end{equation}
\begin{eqnarray}\label{eq54}
\lambda (r,z) = {{{k_0}^2} \over 4}\ln (r) - {k_0}{c_0}{{{r^2}} \over 2} + \frac{1}{4}c_0^2 r^4 \;\;\;\;\;\;\;\;\;\;\;\;\;\;\;\;\;\;\;\;\;\;\\
\;\;\;\;\;\;\;\;\;\;\;\;\;\;\;\;\;\;\;\;\;\;- {\left( {{a_0} + 2{c_0}z} \right)^2}{{{r^2}} \over 2} + {k_0}{a_0}z + {k_0}{c_0}{z^2}&&\nonumber.
\end{eqnarray}
It is important to note that the functions $c(r)=\pm c_0$ and $d(r)=\pm d_0$ in eqs.(\ref{eq25}) and (\ref{eq25a}) turned to be constants $c_0, d_0 <<1$, respectively, once the relativistic effects generated by solar gravity is of the order of $10^{-8}$ times weaker than newtonian ones \citep{yamada}.

In order to obtain an orbit equation to deal with the perihelion advance the related geodesics from Weyl's metric has the components
\begin{eqnarray}\label{eq58}
  {{{{d^2}r} \over {d{s^2}}} + \left( {{\sigma_{,r}} - {\lambda_{,r}}} \right){{\left( {{{dz} \over {ds}}} \right)}^2} + \left( {2{\lambda_{,z}} - 2{\sigma_{,z}}} \right){{dr} \over {ds}}{{dz} \over {ds}}} +    \\ \nonumber
  {{e^{ - 2\lambda }}\left( {{r^2}{\sigma_{,r}} - r} \right)}{{{\left( {{{d\theta } \over {ds}}} \right)}^2} + {e^{4\sigma  - 2\lambda }}{\sigma_{,r}}{{\left( {{{dt} \over {ds}}} \right)}^2}}  & &  \\ \nonumber
  - \left( {{\sigma_{,r}} - {\lambda_{,r}}} \right){{\left( {{{dr} \over {ds}}} \right)}^2} = 0\;, & &
\end{eqnarray}
and also the following set of equations
\vspace{0.5cm}
\begin{equation}\label{eq59}
{2r{\sigma_{,z}}{{d\theta } \over {ds}}{{dz} \over {ds}} - r{{{d^2}\theta } \over {d{s^2}}} + 2r{\sigma_{,r}}{{dr} \over {ds}}{{d\theta } \over {ds}} - 2{{dr} \over {ds}}{{d\theta } \over {ds}} = 0}\;,
\end{equation}
\vspace{0.5cm}

\begin{eqnarray}\label{eq60}
  {{{{d^2}z} \over {d{s^2}}} + \left( {{\lambda_{,z}} - {\sigma_{,z}}} \right){{\left( {{{dz} \over {ds}}} \right)}^2} - \left( {2{\sigma_{,r}} - 2{\lambda_{,r}}} \right){{dr} \over {ds}}{{dz} \over {ds}} +  }& &  \\ \nonumber
  {{e^{ - 2\lambda }}{r^2}{\sigma_{,z}}{{\left( {{{d\theta } \over {ds}}} \right)}^2}+ {e^{4\sigma  - 2\lambda }}{\sigma_{,z}}{{\left( {{{dt} \over {ds}}} \right)}^2}} +  & & \\ \nonumber
   - \left( {{\sigma_{,z}}-{\lambda_{,z}}} \right){{\left( {{{dr} \over {ds}}} \right)}^2} = 0 &&
\end{eqnarray}

\begin{equation}\label{eq61}
{2{\sigma_{,z}}{{dt} \over {ds}}{{dz} \over {ds}} + {{{d^2}t} \over {d{s^2}}} + 2{\sigma_{,r}}{{dr} \over {ds}}{{dt} \over {ds}} = 0}\;.
\end{equation}

As a result, one can obtain the orbit equation
\begin{equation}\label{eq71}
{\left( {{{dr} \over {d\theta }}} \right)^2} = {e^{ - 2\lambda }}\left[ {{r^4}{e^{ - 2\sigma }}\left( {\alpha_0  + \beta_0 {e^{ - 2\sigma }}} \right) - {r^2}} \right]\;,
\end{equation}
where $\alpha_0$ and $\beta_0$ are integration constants.

Due to the structure of Weyl's metric, the coefficient $\sigma$ suffices to produce a nearly newtonian potential from the component $g_{44}$ by the formulae
\begin{equation}
\Phi_{nN}=-\frac{1}{2}(1+g_{44})\;\;,
\end{equation}
reminding that $g_{44}$ metric component is exact and
non-approximated solution, which carries all the non-linear effects. Considering a conformastatic solution \citep{antonio} for eq.(\ref{eq71}) in the sense that $\lambda=0$, which means that the non-linear effects were smoothly attenuated, one obtains
\begin{equation}\label{orbit}
\left(\frac{dr}{d\theta}\right)^2 =\; \left[ r^4 e^{-2\sigma} \left(\alpha_0 + \beta_0 e^{-2\sigma} \right)- r^2 \right]\;,
\end{equation}
which can be transformed into the following conformastatic orbit equation
\begin{equation}\label{orbit1}
\left(\frac{du}{d\theta}\right)^2 + u^2 =\; e^{-2\sigma} \left(\alpha_0 + \beta_0 e^{-2\sigma} \right)\;,
\end{equation}
where the new variable $u$ stands for $u= \frac{1}{r}$ and $\sigma= \sigma(u)$.

As a result, using the method shown in \cite{harko}, eqs. (\ref{eq42}) and (\ref{orbit1}), the perihelion advance is given by
\begin{equation}\label{perih}
\delta\phi = \pi \frac{dF(u)}{du}\rfloor_{u=u_0}\;,
\end{equation}
where a nearly circular orbit is given by the roots of the equation $F(u_0)=u_0$, and the function $F(u)$ is determined by
$$F(u)= \frac{1}{2} \frac{dG(u)}{du}\;,$$
that results the orbit equation in a form
\begin{equation}\label{orbit2}
\left(\frac{du}{d\theta}\right)^2 + u^2 = G(u)\;,
\end{equation}
where $G(u)$ is a generic function that depends on the variable $u$. In the present problem, eq.(\ref{orbit1}) has the form of eq.(\ref{orbit2}).
Moreover, it was shown in \citep{Capistrano} that the appropriate truncated form for the perihelion problem lies in the second order of Weyl's metric expansion, and one can find
\begin{equation}\label{perih5}
    \delta\phi = \delta \phi_{sch} \pm \beta_0 \nu \eta^*\;,
\end{equation}
where we denote $\delta \phi_{sch}=\frac{6\pi GM}{c^2 \gamma(1-\epsilon^2)}$ that stands for standard Einstein's result for Schwarzschild solution, $\gamma$ denotes the semi-major axis and $\epsilon$ denotes the eccentricity of the orbits. The term $\nu$ is the keplerian mean motion given by $\sqrt{\frac{GM}{\gamma^3}}$. The sign $\pm$ refers to the particle deviation to the plane of the orbits and the scale factor $\eta^{*}$ is defined by $\eta^{*}=(180/\pi)(3600)T$, and $T$ is the period of revolution. For the perihelion advance, the term $\beta_0$ was found to be
\begin{equation}\label{perih4}
    \beta_0 =  \epsilon^4 \sqrt{1-\epsilon^2}\;,
\end{equation}
restricted to the interval $[0,1]$ valid for all planets. It is important to note that eq.(\ref{perih5}) can provide both advance and retrograde apsidal precession. In the present framework, as a first approximation, we consider the planets as particles and ignore their rotation. We complement this introduction doing an application to the determination of the theoretical perihelion of giant planets. For calculations, we adopt the Newtonian constant of gravitation $G= 6.67384\times10^{-11} m^3 kg^{-1}.s^{-2}$ \citep{Wilhelm}, one year $1 yr = 365.256 d$, the speed of light $c= 299792458 m/s$ \citep{Wilhelm,Bureau} and the mass of sun $M_{\odot}= 1.98853\times 10^{30} kg$. The relevant physical data are shown in table 01.

In table (02), we present the results of application of eq.(\ref{perih5}). The terms $\delta \phi^{+} $ and $\delta \phi^{-} $ refer to solutions with positive and negative signs. As a matter of completeness, the results are compared to the giant planets perihelion deviation. In this case, we compare not only to GR but also we have chosen, e.g, the Azimuthally Symmetric Theory of Gravitation (ASTG) \cite{nambuya} as a complement, which defines the classical Laplace-Poisson's equation as a cornerstone for understanding local gravitational phenomena. The perihelion advance from ASTG is denoted by $\delta \phi_{ASTG}$ and to GR is denoted by $\delta \phi_{sch}$. Accordingly, it is showing a good agreement with observations, in case of Jupiter and Saturn, and, in the case of Uranus and Neptune, to theoretical  predictions. In the cases presented, we do not observe any relevant difference between the precessions $\delta \phi^{+} $ and $\delta \phi^{-}$ provided by this model. As expected, the parameter $\beta_0$ is also restricted to the interval $[0,1]$ in a agreement with those ranges found in the perihelion of the inner planets and comets in \cite{Capistrano}.
\begin{table*}
  \centering
  \begin{minipage}{140mm}
  \caption{Comparison between the values for secular precession of giant planets in units of arcsec/century of the standard (Einstein) perihelion precession $\delta \phi_{sch}$ \citep{nambuya} and the conformastatic solution $\delta \phi_{model}$. The $\delta \phi_{obs}$ stands for the secular observed perihelion precession in units of arcsec/century adapted from \citep{nambuya} by adding a supplementary precession corrections from EPM2011 \citep{Pitjeva,Pitjev}. The perihelion deviation $\delta \phi_{ASTG}$ denotes the theoretical prediction from ASTG. }
  \begin{tabular}{@{}llrrrrrrr@{}}
  \hline
      Object             &    $\delta \phi_{obs}$   &  $\delta \phi_{sch}$   &$\delta \phi_{ASTG}$   &
   \multicolumn{4}{c}{$\delta \phi_{model}$$\;\;\;\;\;\;\;\;\;$ $\;\;\;\;\;\;\;\;\;$$\beta_0$} & \\
                                                                             &&& $\;\;\;\;\;\;\;$  &$\delta \phi^{+} $& $\delta \phi^{-} $&  &   \\

  \hline
  Jupiter             & $0.070 \pm 0.004$                       &   0.0628     &$0.0700 \pm 0.0200$                   &  $\;\;$0.0622       &  0.0622    & $5.711\times10^{-6}$& \\
                      &                                           &                          &&  \\

  Saturn              & $0.014 \pm 0.002$                       &   0.0138     &$0.0190\pm 0.0005$                   &  $\;\;\;$0.0136      & 0.0136 & $1.017\times10^{-5}$& \\
                      &                                           &                          &&  \\

  Uranus            & \footnote{Not available reliable data.}     &   0.0024     &$0.0025\pm 0.0007$                  & 0.0024               & 0.0024   & $0.4357\times10^{-5}$\\
                      &                                           &                          &&  \\

  Neptune             & \footnote{Not available reliable data.}   &   0.0008     &$0.0027\pm 0.0007$                  & 0.0008               &  0.0008  & $1.630\times10^{-8}$    \\
                      &                                           &                          &&  \\

  \hline
 \end{tabular}
\end{minipage}
\end{table*}

\begin{table*}
  \centering
  \begin{minipage}{140mm}
  \caption{Comparison between the values for anomalous apsidal precession $\delta \phi_{anom}^{a,b,c}$ of the planets in units of miliarcsec/century to the EPM2008 and EPM2011 \citep{Pitjeva,Pitjev} and the INPOP10a planetary ephemerides \citep{Fienga}, respectively.}
  \begin{tabular}{@{}llrrrrrrr@{}}
  \hline
   Object             & $\delta \phi_{anom}^a$  & EPM2008 &$\;\;$ $\delta \phi_{anom}^b$ & EPM2011 & $\delta \phi_{anom}^c$      & INPOP10a   \\

  \hline
  Mercury             & -3.7945                 &$-3.6\pm 5.0$   &   -2.0237            & $-2.0\pm 3.0$          &  0.4047      &$0.4\pm 0.6 $  \\
                      &                         &              &                        &              &&  \\
  Venus               & -0.3878                 &$-0.4\pm 0.5$   &    2.6585            & $2.6\pm 1.6$   &  2.6493              &$0.2\pm 1.5$ \\
                      &                         &              &                        &              &&  \\
  Earth               & -0.1869                 &$-0.2\pm 0.4$   &    1.8694            & $0.19\pm 0.19$  & -0.1669              &$-0.2\pm 0.9$  \\
                      &                         &         &                             &              &&  \\
  Mars                & -0.1057                 &$-0.10\pm 0.5$  &   -0.1939            &$-0.02\pm 0.037$   &  0.3877              &$-0.04\pm 0.15$ \\
                      &                         &         &                             &              &&  \\

  Jupiter             &-------                 &-------        &   60.28                & $58.7\pm 28.3$  &  -48.0000            &$-41\pm 42$ \\
                      &                         &         &                                          &&  \\

  Saturn              & -5.7876                 &$-6.0\pm 2.0$   &   -0.2829            & $-0.32\pm 0.47$  &  0.1910              &$0.15\pm 0.68$ \\
                                                &         &                           &              & &  \\

  \hline
 \end{tabular}
\end{minipage}
\end{table*}

\section{Anomalous precession}
\subsection{The heuristic approach}
Starting from considering the anomalous precession just a fraction of the apsidal perihelion precession advance, from eq.(\ref{perih5}) we can write the following expression
\begin{equation}\label{perih6}
    \delta\phi_{anom} = \pm \frac{\Delta(\delta \phi)}{ \eta^{*}\nu}\;,
\end{equation}
where $\delta\phi_{anom}$ is the anomalous precession and $\Delta(\delta\phi)$ denotes the percentage difference given by $\Delta(\delta\phi)= n \left(\delta\phi_{obs}-\delta\phi_{sch}\right)$. The term $\delta\phi_{obs}$ refers to the perihelion value from observations and $n$ is the percentage number. The results are shown in table (03).

In order to avoid error propagation, we explore the possibility to have a range to vary a percentage fraction, once we have a considerable difference between the studied observational data (e.g, in EPM2013 \citep{Pitjeva2} we do not have extra-precession). Our margin of errors were very small $< 10^{-8}$ and they were omitted here. As it can be observed in table (03), the results are very close to observations as compared one-by-one. As a first prior, from Mercury to Saturn, to model the EPM2008 ephemerides we have the percentage range varying as $0.55 < n(\%)< 0.009 $. Moreover, to EPM2011 ephemerides and INPOP10a, we have the percentage ranges varying as $50 < n(\%)< 0.01$ and $ 6.67< n(\%)< 0.01 $, respectively. This percentage enlargement is not surprising at all since we have a larger margin of errors in Jupiter and Saturn observations. The drawback of this approach, of course, is that it cannot show what is the tendency of the parameter $n$ regarding these three datasets.

\begin{table*}
  \centering
  \begin{minipage}{180mm}
  \caption{Comparison between the values for anomalous apsidal precession $\delta \phi_{anom}^{a,b,c}$ of the planets in units of miliarcsec/century to the EPM2008 and EPM2011 \citep{Pitjeva,Pitjev} and the INPOP10a planetary ephemerides \citep{Fienga}, respectively, using the Chi-squared statistics. The term $<\delta \phi_{anom}>$ denotes the mean-value of $\delta \phi_{anom}^{a,b,c}$ precession values and the p-values are the associated probabilities of the Chi-squared distribution (FIT-P).}
  \begin{tabular}{@{}llrrrrrrrr@{}}
  \hline
   Object             & $\delta \phi_{anom}^a$  & EPM2008 &$\;\;$ $\delta \phi_{anom}^b$ & EPM2011 & $\delta \phi_{anom}^c$      & INPOP10a
   &  $<\delta\phi_{anom}>$ &$\emph{n}$-parameter & FIT-P\\

  \hline
  Mercury             & 0.7340                &$-3.6\pm 5.0$   &   0.5761            & $-2.0\pm 3.0$          &  0.3393      &$0.4\pm 0.6 $ & 0.5498                    &$-0.4052 \pm 0.7599$                   & 0.5401    \\
                      &                         &              &                        &              &&  \\
  Venus               & 0.1841                &$-0.4\pm 0.5$   &    0.1239          & $2.6\pm 1.6$           &  0.1732       &$0.2\pm 1.5$ & 0.1604                    &$ 0.2213 \pm 0.7063$                   & 0.2076   \\
                      &                         &              &                        &              &&  \\
  Earth               & 0.1083                &$-0.2\pm 0.4$   &    0.0978         & $0.19\pm 0.19$         & -0.1083       &$-0.2\pm 0.9$  & 0.1048                    &$-0.1101 \pm 0.1275 $                  & 0.6229  \\
                      &                         &         &                             &              &&  \\
  Mars                & -0.0229               &$-0.10\pm 0.5$  &   -0.0216         &$-0.02\pm 0.037$        & -0.0219    &$-0.04\pm 0.15$ &0.0221                     &$0.0183 \pm 0.00004$                   & 0.9801   \\
                      &                         &         &                             &              &&  \\

  Jupiter             &-------                 &-------         &   364.720             & $58.7\pm 28.3$         &  -254.888      &$-41\pm 42$     &54.916                     &$0.0292 \pm 0.00002$                   & 0.9981   \\
                      &                         &         &                                          &&  \\

  Saturn              & -6.0155                &$-6.0\pm 2.0$   &   -0.3334           & $-0.32\pm 0.47$  &  0.1364           &$0.15\pm 0.68$ & -2.0731             & $0.0047 \pm 0.00003$    & 0.9994    \\
                      &         &                           &              & &  \\
  \hline
 \end{tabular}
\end{minipage}
\end{table*}

\subsection{The Levenberg-Marquardt algorithm}
To get information on the tendency of the parameter \emph{n} (hereon, not interpreted as a percentage number) based on the three datasets to control the systematics, we use GnuPlot 5.0 software to compute non-linear least-squares fitting applying the Levenberg-Marquardt algorithm.

Due to the fact that the observations have high error bars and the difficulty to obtain fair estimates of the errors on many observational quantities, the analysis of the Chi-squared values can only tell us relative merits of different models in this case, which makes the error distribution not representable as a random error in the standard statistics and perhaps it may lead to the appearance of eventual outliers.

On the other hand, the associated probability $p$ values, which measure the data and the model compatibilities, may be of use.  These values indicate that if $p > 0.98$ we have a good data plus model fit and $p < 0.02$ means an inconclusive analysis systematic effects or a new model is needed. The ``average'' fit is obtained from a range of $0.02 < p < 0.98$. From Mercury to Earth, we note that the probability achieves the ``average'' fit and from Mars to Saturn, the high error bars seems to dominate the systematics with high probability  and very small values for the reduced Chi-squared close to zero.

On the contrary what happened to the previous subsection with the heuristic approach, interestingly, we obtain a tendency of the parameter driven by the data. Except for Jupiter, which only its mean-value perihelion $<\delta \phi_{anom}^{Jupiter}>$ is compatible with EPM2011, the other planets have a tendency to approximate to the INPOP10a planets ephemerides data. In this case, it leads to the conclusion that the observed anomalies are statistically irrelevant \citep{Acedo,lorio14, Fienga}.

\section{Final remarks}
In the present paper, we have studied the Weyl conformastatic solution in the nearly-newtonian regime. This gravitational regime was originally proposed and qualitatively discussed in \citep{wheeler} as an intermediate regime somewhere in between newtonian gravity and GR. What makes different from the well-known standard PPN approximation is the possibility that we can study more the influence of non-linear effects, since the nearly-newtonian regime is obtained only from the geodesic equations, the non-linearities of the gravitational field, in principle, are unspoiled, and may be ``tuned'' to a specific end.

The main advantages of this method are twofold. First, since the diffeomorphic invariance from GR is lost, the choice of an appropriate geometry to a specific problem becomes an important matter. This seems to be appropriate for astrophysics purposes, once today's sophisticated astrophysical observations are made in a three-dimensional space, restating the kantian relativity of forms, i.e, the geometry is the representation of the real forms of the world. Secondly, it resides in its simplicity since the modification of Einstein gravity is not necessary nor the inclusion of exotic terms.

In this sense, we have applied the Weyl conformastatic vacuum solution considering the planets as an orbiting particle in a thin disk expansion. From the perihelion formulae obtained in a previous publication \citep{Capistrano}, that predicts both advance and retrograde precessions, we have tested this methodology using only one parameter related to the eccentricity of the orbits restricted to the interval $[0,1]$ determining the perihelion precession to giant planets obtaining a close agreement to observations and the specialized literature. The main test we wanted to provide was to submit the methodology to study the anomalous perihelion of planets.

Initially, we have applied a heuristic approach in order to get a first insight on the problem and to understand how the parameter would behave and how the systematic errors would inflict on the results. Surprisingly, comparing the results to observations one-by-one we have obtained a well-accommodated results. However, due to the high error bars, it was necessary to understand the tendency of the fitting parameter would be driven and the heuristic approach would not suffice. In order to solve this matter, we have applied the Levenberg-Marquardt algorithm and the related Chi-squared statistic. We have shown that the fitting parameter has tended to close agreement with the INPOP10a planetary ephemerides results. The state-of-the-art suggests that the anomalous perihelion is statistically irrelevant hence disproving a new eventual effect, which is compatible with our results. As we are trying to show that this nearly newtonian limit can be an interesting arena to investigate in solar system physics (only controlling the strength of the gravitational field by the deviation equation and leaving the geodesics intact), at first sight, we do not expect any improvement from a modified gravitational theory (with general relativity as its limit) without improvement on the level of confidence of observations.

A final definition might be achieved in the near future improvement in accuracy from astronomical/astrophysical observations with new experimental techniques and spacial missions, e.g, the Uranus Pathfinder123, Outer Solar System-OSS124 and the New Horizons spacecraft missions. We conjecture that a convincingly precision for this phenomenon, if exists, might be achieved at arcsec/century level. As future perspectives, an extended analysis to nodal precession in spheroidal metrics are currently in due course.

\label{lastpage}

\end{document}